\documentclass[aps,pra,twocolumn,superscriptaddress]{revtex4}
\usepackage{amssymb}
\usepackage{graphicx}% Include figure files

\newcommand{\ds}{\displaystyle}

\newcommand{\be}{\begin{equation}}
\newcommand{\ee}{\end{equation}}

\newcommand{\w}{\omega}
\newcommand{\W}{\Omega}
\newcommand{\g}{\gamma}
\newcommand{\G}{\Gamma}

\newcommand{\bra}{\langle}
\newcommand{\ket}{\rangle}

\begin{document}

\title{Optical imaging beyond the diffraction limit via dark states}

\author{Hebin Li}
\affiliation{
    Department of Physics and Institute for Quantum Studies,
    Texas A\&M University,
    College Station, Texas 77843-4242
}
\author{Vladimir A. Sautenkov}
\affiliation{
    Department of Physics and Institute for Quantum Studies,
    Texas A\&M University,
    College Station, Texas 77843-4242
}
\affiliation{
P.N. Levedev Physical Institute, 53 Leninsky prospect, Moscow 119991,
Russia
}
\author{Michael M. Kash}
\affiliation{
    Department of Physics and Institute for Quantum Studies,
    Texas A\&M University,
    College Station, Texas 77843-4242
}
\affiliation{
Department of Physics, Lake Forest College, Lake Forest, Illinois 60045
}
\author{Alexei V. Sokolov}
\affiliation{
    Department of Physics and Institute for Quantum Studies,
    Texas A\&M University,
    College Station, Texas 77843-4242
}
\author{George~R.~Welch}
\affiliation{
    Department of Physics and Institute for Quantum Studies,
    Texas A\&M University,
    College Station, Texas 77843-4242
}
\author{Yuri V.\ Rostovtsev}
\affiliation{
    Department of Physics and Institute for Quantum Studies,
    Texas A\&M University,
    College Station, Texas 77843-4242
}
\author{M. Suhail Zubairy}
\affiliation{
    Department of Physics and Institute for Quantum Studies,
    Texas A\&M University,
    College Station, Texas 77843-4242
}
\affiliation{
    Texas A\&M University at Qatar,
    Education City, P.O. Box 23874, Doha, Qatar
}
\author{Marlan O. Scully}
\affiliation{
    Department of Physics and Institute for Quantum Studies,
    Texas A\&M University,
    College Station, Texas 77843-4242
}
\affiliation{
Applied Physics and Materials Science Group, Engineering Quad, Princeton
Univ., Princeton 08544
}

\date{\today}

\begin{abstract}
We study the possibility of creating
spatial patterns having subwavelength size by
using the so-called dark states formed by the interaction
between atoms and optical fields. These optical fields
have a specified spatial distribution.
Our experiments in Rb vapor display spatial patterns that are smaller
than the length determined by the diffraction limit of the optical system used
in the experiment.
This approach may have applications to interference lithography
and might be used in coherent Raman spectroscopy to create patterns with
subwavelength spatial resolution.
\end{abstract} %}}}

\maketitle

%\section{Introduction}

%
\begin{figure}[b] %1
\center{
\includegraphics[width=6cm,angle=90]{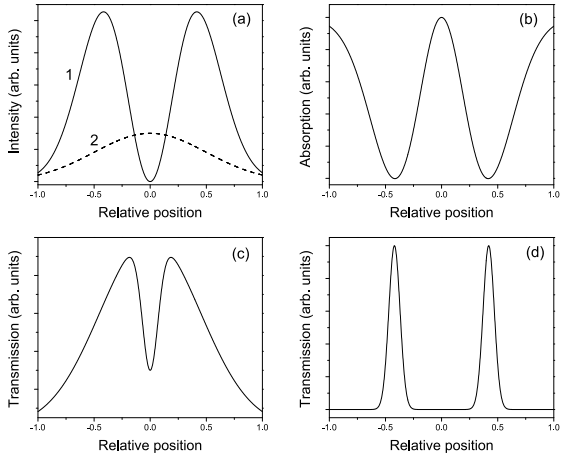}
}

\caption{\label{fields}
Qualitative description of the idea.
(a) Distribution of the drive (1) and the probe (2) fields vs
a transverse spatial coordinate at the entrance to the cell.
(b) Dependance of the absorption coefficient given by Eq.(\ref{kappa})
vs position.
Plots (c) and (d) show the distribution of the probe beam
after propagating through the
cell. Case (c) is for a strong drive field and relatively low optical density.
Case (d) is for a relatively weak drive field and large optical density.
}
\end{figure}

The ability to create small images is important for material processing
technology and for improving the resolution of microscopy
for bio-medical applications~\cite{lithography}.
Recently, several methods have been presented that are able to overcome the
diffraction limit of the imaging system. Quantum microscopy is based on using
a nonclassical optical field approach~\cite{scully,dowling}.
Microscopy with classical fields can be enhanced by the nonlinear
optical response of the medium~\cite{hells}.
Classical field amplitude and phase arrangements can be used to locate
the position of an atom with subwavelength precision in an atomic beam
\cite{thomas,thomas2,thomas3,qamar00pra} and in a cavity~\cite{zubairy}.

Here, we suggest a new approach that is based on coherent population
trapping~\cite{i-eit,arimondo96,book,marangos98jmo,eit05rmp}.
Optical fields applied to a three-level quantum system
excite the so-called {\it dark state}, which is decoupled from
the fields.
Similar approaches using coherent population
trapping have also been developed by several groups
(for example, see~\cite{zubairy08prl,lukin,yavus,agarwal}).

\begin{figure}[htb]
\includegraphics[width=1\columnwidth]{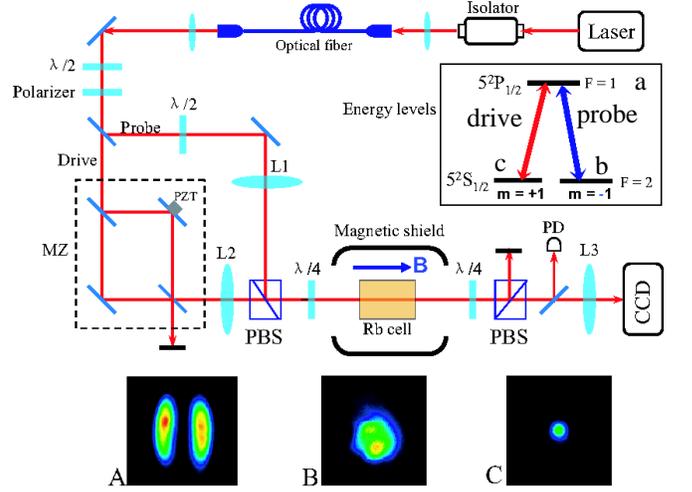}% experimental result
\caption{\label{fig1}(Color online) Experimental schematic. $\lambda$/2:
half-wave plate; $\lambda$/4: quarter-wave plate; L1, L2, L3: lenses;
MZ: Mach-Zehnder interferometer; PZT: piezo-electric element;
PBS: polarizing beam splitter, PD: photo diode; CCD: CCD camera.
Picture A is the spatial
intensity distribution of the drive field. Picture B is the
beam profile of the parallel probe beam without the lens L1. Picture C
is the beam profile of the diffraction limited probe beam with the lens L1.
The inset is the energy diagram of the Rb atom, showing representative
sublevels.}
\end{figure}

As a qualitative introduction,
assume that the drive field Rabi frequency $\W_d$ has the particular spatial
distribution sketched in Fig.~\ref{fields}(a) by the solid line (1).
The weak probe field Rabi frequency $\W_p$ ($\W_p\ll \W_d$) has a diffraction
limited distribution (shown by the dashed line (2) in Fig.~\ref{fields}(a)).
The probe and drive fields are applied to the atom (see the inset in
Fig.~\ref{fig1}). At all positions of nonzero drive field, the dark state,
which is given~\cite{book} by
$|D\ket=(\W_p|c\ket - \W_d|b\ket)/\sqrt{\W_p^2 + \W_d^2}$,
is practically $|b\ket$. When the drive field is zero, the dark state
is $|c\ket$, and the atoms at these positions are coupled to the fields and
some atoms are in the upper state $|a\ket$. The size of a spot where the
atoms are excited depends on the relaxation rate $\g_{cb}$
between levels $|b\ket$ and $|c\ket$.
For $\g_{cb}=0$, the size of spot is zero, smaller than the optical wavelength.

The Hamiltonian of three-level atom
interacting with optical fields (see the inset in Fig.~\ref{fig1})
is given by
\be
H = \hbar\W_d|a\ket\bra b| + \hbar\W_p|a\ket\bra c| + adj.,
\ee
where $\W_{p,d} = \wp_{p,d} E_{p,d}/\hbar$ are the Rabi frequencies of the
drive $E_d$ and the probe $E_p$ fields, respectively. Then, the atomic response
is given by the set of density matrix equations~\cite{book}
\be
\dot\rho = -{i\over\hbar}[H,\rho] -{\G\rho + \rho\G\over 2}
\ee
where $\G$ describes the relaxation processes. The propagation of the probe field $\W_p$
through the cell is governed by Maxwell's Equations and, for propagation in
the $z$-direction, can be written in terms of the probe field Rabi frequency as
\be
{\partial\W_p\over\partial z} = - i\eta\rho_{ab}
-i{1\over 2 k}{\partial^2\over\partial x^2}\W_p.
\label{propagation}
\ee
The first term accounts for the dispersion and absorption of the
resonant three-level medium, and the second term describes the focusing and/or
diffraction of the probe beam. The density matrix element $\rho_{ab}$
is related to the probe field absorption which in turn depends on the detuning
and the drive field. This is characterized by an absorption coefficient:
\be
\kappa = \eta{\G_{cb}\over\G_{ab}\G_{cb} +|\W_d(z,x)|^2},
\label{kappa}
\ee
where $\G_{cb} = \g_{cb} + i\w$ and $\G_{ab} = \g+i\w$;
$\w=\w_{ab}-\nu$ is the detuning from the atomic frequency $\w_{ab}$;
$\g$ is the relaxation rate at the optical transition;
and $\eta = 3\lambda^2 N\g_r/8\pi$; $\g_r$ is the spontaneous emission rate.
We now assume that the drive field has
a distribution of intensity near its extrema given by
\be
|\W_d(z,x)|^2 = |\W_0|^2
\left\{
\begin{array}{cc}
\left[1 - \left({x-x_0\over L}\right)^2\right], & \mbox{$x \simeq x_0$,} \\
\left({x\over L}\right)^2, & \mbox{$x\ll L$},
\end{array}
\right.
\label{Wd}
\ee
where $\W_0=\W_d(z,x_0)$,
a typical absorption profile vs $x$ is shown in Fig.~\ref{fields}(b).
Neglecting the diffraction term in Eq.(\ref{propagation}), we can write
an approximate solution for Eq.~(\ref{propagation}) as
\be
\W_p(z, x) = \W_p(z=0, x)\exp(-\kappa z).
\ee

For relatively low optical density ($\kappa z \simeq 1$),
nearly all of the probe field propagates through the cell except for
a small part
where the drive field is zero (see Fig.~\ref{fields}(a)).
Absorption occurs there because the probe beam excites
the atomic medium. The width of the region of the excited medium, in the
vicinity of zero drive field, is characterized by
\be
\Delta x = L\ds\sqrt{\G_{ab}\G_{cb}\over|\W|^2},
\ee
where $\W=\W_d(z=0,x=0)$. This region is small, but its contrast is limited
because of the finite absorption of the medium at the center of optical line
(Fig.~\ref{fields}(c)).

For higher optical density, this narrow feature becomes broadened
(compare Fig.~\ref{fields}c and d), but two narrow peaks are formed during
the propagation of the probe beam (see Fig.~\ref{fields}(d)). For zero
detuning, their width is given by
\be
\Delta x = L\sqrt{|\W|^2\over\eta\g_{cb}z}.
\ee

Note that the propagation occurs in a waveguide formed by the coherent
medium.  The drive field provides flexibility for creating patterns with
sizes smaller than the wavelength of the laser.
The distribution of fields is governed by electrodynamics and has
a diffraction limit, while the distribution of molecules in their excited states
is NOT related to the diffraction limit, but rather determined by the relaxation
rates $\G_{ab}$ and $\G_{cb}$, and thus can have spatial sizes smaller
than the wavelength.

%\section{Experiment}

\begin{figure}[htb]
\includegraphics[width=.9\columnwidth]{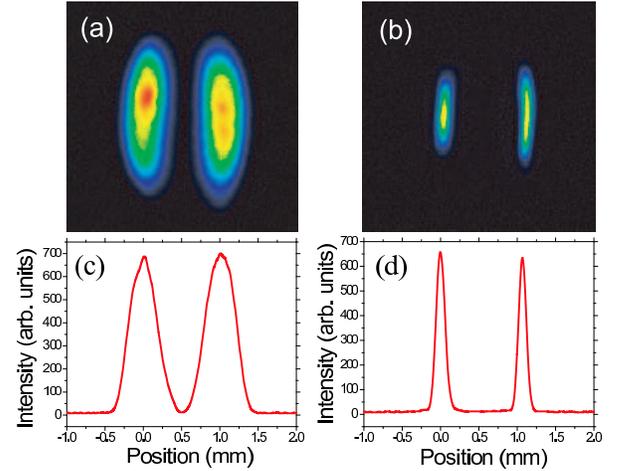}% experimental result
\caption{\label{fig2}(Color online) The results of the experiment
with a parallel probe beam. Picture (a) shows
the image of the intensity distribution
of the drive field in the Rb cell. Picture (b) shows the intensity
distribution of the transmitted probe field. Curves (c) and (d) are
the corresponding profiles. The widths of the peaks in curves (c) and
(d) are 0.4 mm and 0.1 mm, respectively.}
\end{figure}

In this Letter, we report a proof-of-principle experiment
in Rb vapor to demonstrate our approach.
We have observed that the distribution of the transmitted probe beam
intensity has a double-peak pattern, which is similar to that of
the drive beam,
but the width of peaks of the probe beam is narrower than that of the drive
beam.

The experimental schematic is shown in~Fig.~\ref{fig1}.
To obtain diffraction limited focusing the laser beam must have a good quality
spatial profile.
This is obtained by sending the radiation of an external cavity diode laser
through a polarization-preserving single mode optical fiber.
The laser beam is vertically polarized and split into
two beams (drive and probe). The probe beam carries a small portion
of the laser intensity, and its polarization is rotated to be horizontal.

To create a double-peak spatial distribution for the drive field,
the drive beam is split into two beams that cross at a small angle,
using a Mach-Zehnder interferometer (shown in
the dashed square of Fig.~\ref{fig1}).
A typical two-peak interference pattern of
crossing beams is shown as Fig.~\ref{fig1}A.

The probe and drive beams combine on a polarizing beam splitter,
arranged so that the probe field and the interference
pattern of the drive field are overlapped in a Rb cell.
The Rb cell has a length of 4 cm, and is filled with $^{87}$Rb.
A magnetic shield is used to isolate the cell from any environmental magnetic
fields, while a solenoid provides an adjustable, longitude magnetic field. The
cell is installed in an oven that heats the cell to reach an
atomic density of 10$^{12}$~cm$^{-3}$. The laser is tuned to the D$_1$
line of $^{87}$Rb at the transition 5$^2$S$_{1/2}$($F=2$)
$\rightarrow$ 5$^2$P$_{1/2}$($F=1$).

As stated above. the probe and drive beams have the orthogonal linear
polarizations. A quarter-wave plate converts them into
left and right circularly polarized beams, which couple
two Zeeman sublevels of the lower level and one sublevel of the excited level
of the Rb atoms (see the inset of Fig.~\ref{fig1}).

After passing through the cell, the probe and drive beams are converted back
to linear polarizations  by another quarter-wave plate and the separated by a
polarizing beam splitter (PBS). The power of transmitted probe
field is monitored by a photodiode (PD). The spatial intensity
distribution of probe field is recorded by an imaging system,
consisting of the lens L3 and a CCD camera.

The intensity of the probe beam is low enough that its
transmission through the cell is almost zero without the presence of drive
laser.
Applying the drive laser makes the atomic medium transparent for the probe
laser wherever the EIT condition is satisfied. If the drive laser
has a certain transverse spatial distribution, then that pattern
can be projected to the transmission profile of the probe laser.

Two different experiments have been performed. In the first experiment,
the lenses L1 and L2 are not used, and the probe beam is a parallel
beam with a diameter of 1.4~mm. The image of the drive intensity
distribution in
the cell is shown in Fig.~\ref{fig2}(a). The probe intensity has a Gaussian
distribution before entering the cell, and its distribution is similar to the
drive intensity distribution after the cell. As shown in
Fig.~\ref{fig2}(b), however,
the transmitted probe intensity has a distribution that has
sharper peaks compared with the pattern of the drive intensity.
The horizontal cross-sections of the drive and the transmitted probe
distributions are shown in Fig.~\ref{fig2}(c) and (d) respectively.
In the drive intensity profile, the width (FWHM) of the peaks is 0.4~mm.
The width
(FWHM) of the peaks in the transmitted probe intensity profile is 0.1~mm.
The spacing between two peaks are the same for both the drive and
transmitted probe fields. We define the {\it finesse} as the ratio of the
spacing between peaks to the width of peaks. The finesse of
the transmitted probe intensity distribution is a factor of 4 smaller than
that of the drive intensity distribution.

\begin{figure}[htb]
\includegraphics[width=.9\columnwidth]{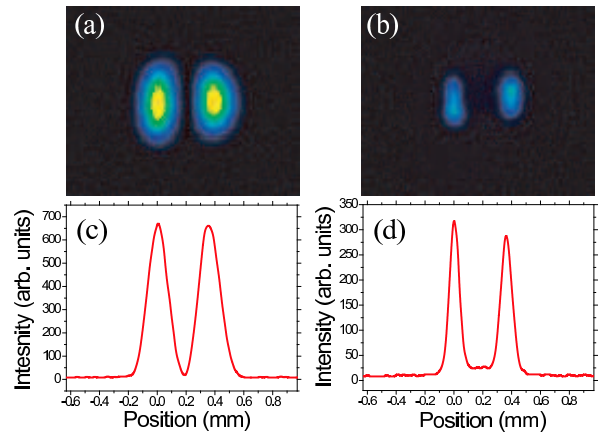}% experimental result
\caption{\label{fig3}(Color online) The results of the experiment
with the diffraction limited probe beam. Picture (a) shows the image of
the intensity
distribution of the drive field in the Rb cell. Picture (b) shows the image of
the intensity distribution of the transmitted probe field. Curves (c) and
(d) are the corresponding profiles. The widths of the peaks in curves
(c) and (d) are 165 $\mu$m and 93 $\mu$m, respectively.}
\end{figure}

\begin{figure}[tb] %1
\center{
\includegraphics[width=6cm]{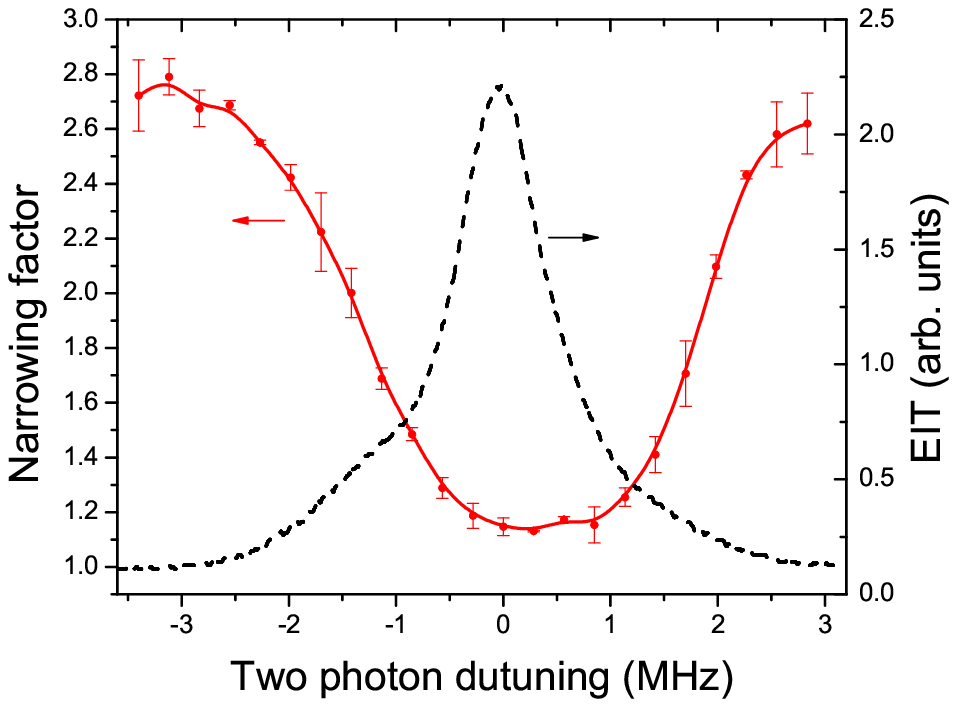}

\includegraphics[width=6cm]{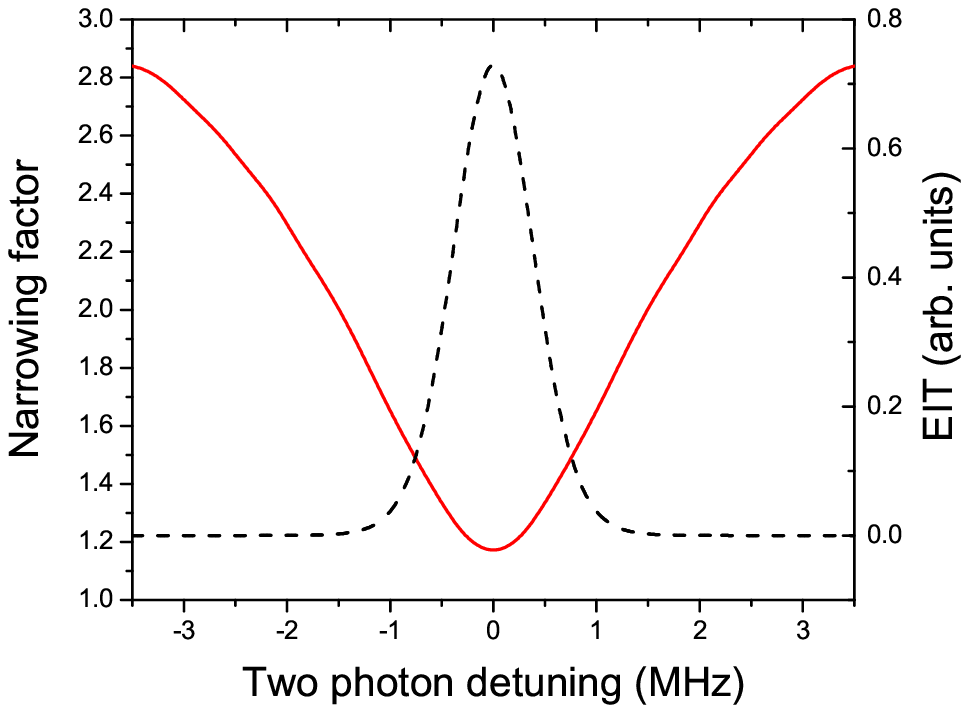}
}
\caption{\label{ratio}
Narrowing of the transmitted probe intensity distribution as function of the
probe detuning:
(a) experimental results (a) and (b) theoretical simulation.
The transmitted probe profile is shown as well.
%\label{fields}
%
}
\end{figure}

In the second experiment, the lenses L1 and L2 are used. A
parallel probe beam (Fig.~\ref{fig1}B) with a diameter of 1.4 mm is focused by
the lens L1, which has a focal length of 750~mm. The beam size at the waist is
0.5 mm, which is diffraction limited. To assure experimentally that the beam is
diffraction limited, we increased the beam diameter of the parallel
beam by the factor of 2, and the beam size at the waist became two
times smaller. The lens L2 is used to make the drive beam smaller
(but not diffraction limited) in the Rb cell, where the pattern of
drive field is spatially overlapped with the waist of the probe
beam. Classically, there should be no structures at the waist of the
probe beam because it is diffraction limited. Structures
can be created in a region smaller than the diffraction
limit in our experiment, however. The
experimental result is shown in Fig.~\ref{fig3}. The drive field still has a
double peak intensity distribution (Fig.~\ref{fig3}(a)). The transmission of the
diffraction limited probe beam also has a double-peak intensity distribution as
shown in Fig.~\ref{fig3}(b). Curves (c) and (d) are the beam profiles of the
drive and transmitted probe beams respectively. The width of the
peaks in the drive beam is 165~$\mu$m, and the width of the peaks
in the transmitted probe beam is 93~$\mu$m. The finesse
of the transmitted probe beam is 1.8 times greater that that of the drive beam.
For the probe beam, the structure created within the diffraction limit has
a size characterized by the width of peaks~(93~$\mu$m). This
characteristic size is 5 times smaller than the size of
the diffraction limited probe beam~(500~$\mu$m).

%\section{Discussion}

Thus, we have demonstrated that our concept works in Rb vapor.
The typical image is shown in Fig.~\ref{fig1}.
One can see that the width of the probe image (C)
is at least three times smaller
than the width of the drive image (A).
Although the diffraction limit is ``beaten,'' the experiment does
not violate any laws of optics. The probe beam is diffraction
limited, but the atoms are much smaller than the size of
diffraction limited beam.
%
% The intensity distribution of the drive beam
% makes atoms at specific locations (where drive laser is strong
% enough) stay in dark states, in which the atoms are transparent to
% the probe field. The transmitted probe beam picks up the pattern of
% the drive beam.
%
Moreover, due to the strong nonlinearity of the EIT,
the characteristic size of the pattern in the
transmitted probe beam is much smaller than that of the drive beam
and the diffraction limit of the probe beam.

We have also measured the narrowing effect vs the detuning of the probe field
and have performed simulations using the density matrix approach.
The results are shown in Fig.~\ref{ratio}.
The calculations reproduce the data satisfactorily.
The dependence on detuning has not been
considered in~\cite{zubairy08prl,lukin,yavus,agarwal}.
It is unique for our approach and can be understood in the following way.
Absorption by the atomic medium given by Eq.(\ref{kappa}) with a drive
intensity distribution given by Eq.(\ref{Wd}) can be written as
\be
\kappa = \eta\left({\g_{cb}\over|\W|^2} + {\g\w^2\over|\W|^4} +
\left({\g_{cb}\over|\W|^2} + 2{\g\w^2\over|\W|^4}\right)\left({x\over L}\right)^2
\right).
\ee
Then, ratio of the width of the probe intensity distribution
to the width of the
drive intensity distribution is given by
\be
R = {L\over\Delta x} = \sqrt{\eta z
\left({\g_{cb}\over|\W|^2} + 2{\g\w^2\over|\W|^4}
\right)}.
\ee
From this we see that the finesse increases with the detuning.

In conclusion, we have performed a proof-of-principle experiment that our
concept works in Rb vapor and have experimentally demonstrated the possibility
of creating structures having widths smaller than those determined
by the diffraction limits of the optical systems.
The results obtained here can be viewed as an experimental verification of our
approach, as well as evidence supporting the theoretical predictions
and results obtained by others~\cite{zubairy08prl,lukin,yavus,agarwal}.
This technique might be used in microscopy by studying the distribution of
molecules with subwavelength resolution or in lithography by manipulating
molecules in the excited state.
Also, note that it may be possible to apply this approach to
coherent Raman scattering (for example, CARS). This may improve
the spatial resolution of CARS microscopy.

We thank P. Hemmer and O. Kocharovskaya %, A. Sokolov
for useful and fruitful discussions, and
gratefully acknowledge the support of the Office of Naval Research and
the Robert A.\ Welch Foundation (Grant \#A1261).

\end{document}